**Light-assisted, templated self-assembly using a photonic-crystal slab**

*Eric Jaquay, Luis Javier Martínez, Camilo A. Mejia, and Michelle L. Povinelli\**

*University of Southern California, Los Angeles, CA 90089*

*\*povinell@usc.edu*

The force of light on objects provides tremendous flexibility for nanoscale manipulation. While conventional optical tweezers use the optical gradient force of a focused laser,[1-3] recent work has leveraged the strong field gradients near microphotonic devices for particle trapping.[4-15] However, such work has focused on trapping single or few particles. We have proposed to use optical forces near microphotonic devices for a fundamentally different purpose: to assemble periodic arrays of nanoparticles resembling synthetic, reconfigurable 2D crystals.[16, 17] Our approach, called light-assisted, templated self-assembly (LATS), exploits photonic-crystal slabs to create resonantly-enhanced optical forces orders of magnitude larger than radiation pressure.[16, 17] Here we provide the first experimental demonstration of LATS, assembling a square array of over 100 polystyrene particles near a silicon photonic-crystal slab. Our method, ideally suited for on-chip integration, should provide a platform for flow-through, serial fabrication of 2D or 3D-nanostructured materials, all-optically tunable photonic devices, and lab-on-a-chip applications.

The LATS process is shown schematically in Figure 1. Light is incident from below on a photonic-crystal slab, which consists of a silicon device layer patterned with a periodic array of air holes. The slab is designed to support guided-resonance modes, electromagnetic modes for which the light intensity near the slab is strongly enhanced.[18] Our previous work has theoretically predicted[16] that when the incident laser is tuned to the wavelength of a guided-resonance mode, nanoparticles will be attracted toward the slab. The attractive, optical force arises from a strong electric-field gradient just above the slab surface. In addition, the nanoparticles will experience lateral optical forces due to the electromagnetic field structure of the guided-resonance mode, resulting in the assembly of a nanoparticle array.

Unlike traditional colloidal self-assembly, for which free energy minimization results in hexagonal, close-packed structures, our process is not subject to such constraints. In this paper, we experimentally demonstrate the formation of a square lattice as one such example. Indeed, the use of light to drive the system dramatically alters the underlying potential landscape, potentially allowing for the formation of a range of complex lattices[16, 17] and multiparticle clusters.[19] Nanoparticle arrays assembled using LATS can be viewed as "programmable optical matter:"[20] turning the laser on and off will reversibly assemble or disassemble the structure. Moreover, exciting different resonance modes of the photonic crystal, by adjusting the wavelength of the input laser, should allow different crystalline structures to be formed.

Light-driven assembly of multiparticle patterns has previously been achieved using structured light fields generated by interference fringes, holography, spatial light modulators, or other methods.[21-23] Our approach differs crucially from previous work in that it exploits near-field, rather than far-field, effects. Rather than generating a structured light beam via free-space optics, we use a simple, Gaussian input beam. The structured light field responsible for trapping is generated by the interaction of light with the photonic-crystal device. Here, we demonstrate the method using an external laser incident on a photonic-crystal slab. Ultimately, however, LATS could be carried out using a photonic-crystal laser, allowing the integration of the light source with the trapping device and making our approach highly suitable for on-chip integration. We thus expect a wide range of applications from all-optically tunable photonic devices, to materials assembly, to biological trapping and manipulation.

**Results**

We designed, fabricated, and characterized a photonic-crystal slab for use in the LATS process. An electron micrograph is shown in Figure 2(a). The device was fabricated in silicon using electron-beam lithography and reactive ion etching (see Methods). The dimension and spacing of the holes was designed to support doubly-degenerate guided-resonance modes near 1.55 μm. The magnetic field profile resulting from an *x*-polarized, incident plane wave is shown in Figure 2(b). Fields were calculated using the three-dimensional finite-difference time-domain method (FDTD). The field-profile for a *y*-polarized incident wave is rotated by 90 degrees. Figure 2(c) shows the measured transmission spectrum of the device. The guided-resonance mode appears as a dip in the spectrum. The quality factor $Q$ was determined to be ~170 by fitting to a Fano-resonance shape.

We carried out the assembly process in a microfluidic chamber filled with 520-nm diameter polystyrene particles, using a laser power of 64 mW. Figure 3 shows snapshots of the LATS process. The photonic-crystal lattice is visible in the background of each frame. When the laser beam is turned on, nanoparticles are attracted to the slab, and begin to occupy sites of the square lattice (Figure 3(a)). As time progresses, additional particles diffuse into the region where the beam intensity is high, and begin to form a cluster (Figure 3(b)). Eventually, a regular array of particles is formed (Figure 3(c)). The square symmetry of the assembled particles is evident from the picture. When the laser beam is turned off, the particles immediately begin to disperse and diffuse away from the slab (Figure 3(d)). Full videos of the assembly and release processes are included in the Supplementary Information. The frames in Figure 3 were recorded with a dilute particle solution for clarity of imaging, and represent an elapsed time of approximately one hour. Faster cluster formation occurs with solutions of higher concentration.

Each site of the square lattice may be viewed as an optical trap. We used particle-tracking software to analyze particle motion for fully-assembled clusters (see Methods). Figure 4(a) shows the recorded particle positions extracted from a 20-second video. The incident light was polarized along the $x$-direction of the lattice. The figure shows that the particles tend to stay above the holes in the photonic crystal, with some variation in position over time. Each blue ellipse represents a fit to the data in a single unit cell. It can be seen that the variation in particle position increases at the edge of the trapping region, due to the reduction in power away from the center of the beam.

The stiffness of each trap can be determined from the variance in particle position.[24] Figures 4(b) and 4(c) show histograms of the in-plane stiffness values extracted from the videos. The stiffness of each trap was normalized to the local intensity in that unit cell (see Methods). We observe that the power-normalized stiffness over the array of traps is normally distributed, both for the parallel and perpendicular stiffness. The mean parallel stiffness is lower than the perpendicular stiffness, as shown in Table 1 (0° angle).

We observe that the trap stiffness can be tuned by rotating the direction of incident light. When the incident light is polarized at 45° with respect to the lattice directions, the stiffness values are approximately the same in the $x$- and $y$-directions (Table 1). This is to be expected, since the incident light excites both of the doubly-degenerate modes with equal strength. At 90°, the stiffness in the parallel direction ($y$) is again lower than in the perpendicular ($x$) direction. The ability to tune the stiffness with incident light indicates the strong optical nature of our traps. The mean values of stiffness are comparable to those reported elsewhere in the literature for single particle traps.[25]

Using the Stokes drag method (see Methods), we experimentally estimated the maximum force exerted on the particles by the traps to be 0.3 pN.

To understand how the optical forces result in the observed nanoparticle patterns, we calculated the force numerically (see Methods). Figure 5(a) shows the force on a 520-nm diameter particle whose bottom edge is 25 nm above the surface of the photonic crystal slab for *x*-polarized light. The background color represents the vertical force, where a negative force indicates attraction toward the slab. There are two regions above and below the hole where the force is slightly repulsive, but at any other position within the unit cell the particle is attracted to the slab. The arrows represent the in-plane forces.

To determine the equilibrium position of the trapped particle, we calculate the optical potential. Given the relative size of the particles (260 nm radius) and holes (150 nm radius) in our experiment, the particle can be partially drawn into the hole by the attractive vertical force. We calculate the optical potential as a function of *x-y* position. For each *x-y* position, the vertical height of the particle is as small as possible, given the geometrical constraints (see inset in Figure 5(b)). The result is shown by the blue line ('total potential') in Figures 5(b) and (c). It can be seen that the stable equilibrium position is in the center of the hole ($x = 0$, $y = 0$), in agreement with experiments.

The ability of the particle to sink into the hole is a key factor in determining the equilibrium positions. For comparison, the red, dashed line ('in-plane potential') in Figure 5(b) shows the optical potential calculated at a constant *z*-height (bottom edge of particle 25 nm above the slab surface). Two local minima are observed at the edges of the hole, which are indicated by green, dashed lines. From inspection of Figure 5(a), it can be seen that these points correspond to locations where the in-plane forces are zero. However, these two minima are not stable equilibrium positions of the total potential (Figure 5(b)).

From our experiments, we determined that the threshold intensity for trapping was 134 µW per unit cell (see Methods). For this intensity, the calculated potential depth is 4.5 $k_B T$.

**Discussion**

In summary, we have experimentally demonstrated the technique of light-assisted, templated self-assembly (LATS). Our technique uses the resonantly enhanced near field of a photonic-crystal slab to create periodically spaced optical traps. Nanoparticles in solution are attracted to the slab and form ordered arrays. We have observed the assembly of 100 polystyrene particles in a square lattice using 64 mW of incident power. We have measured the trapping stiffness as a function of incident polarization, and we have shown that the equilibrium trapping positions can be predicted via calculation of the optical forces.

Our technique can be extended to assemble larger numbers of particles by designing the photonic-crystal slab to reduce the input power per area required for trapping. The input beam can then be spread over a larger area, resulting in a larger cluster. One approach to reducing the required power per area is to use a mode with a higher quality factor, resulting in higher near-field intensity. Another approach is to use slot-confinement effects to strongly localize the field in the trapping regions, as we have previously studied theoretically.[17]

The LATS approach can be used to assemble complex structures with symmetries not constrained by the typical free energetic constraints; here we have demonstrated just one. We envision that judicious template design will allow the assembly of a variety of lattice types with complex unit cells. Moreover, changing the wavelength of the incident beam to excite a different resonance can be used to reconfigure the particle arrangement.[14] The use of metal nanoparticles or quantum dots is also an area for future exploration.

We anticipate that our technique will find applications in the fabrication of metamaterials and other photonic devices. For example, post-assembly polymerization could be used to transfer 2D nanoparticle arrays to another substrate. Our preliminary results indicate that it is also possible to assemble 3D arrays *in situ*.

LATS should also allow a variety of dynamic, real-time applications. One example is the use of the assembled, reconfigurable "photonic matter" as an all-optically tunable transmission filter. Other applications, based on particle dynamics in the 2D optical potential, include particle sorting[26,27] and ratchet behavior.[28] We also expect that LATS may be extended for batch processing of biological objects, providing a novel tool for reconfigurable control over spatially-mediated biological interactions.

Finally, LATS naturally lends itself to compact integration on-chip. By fabricating the photonic-crystal device in an active material, the light source could be integrated with the trapping device, allowing for system miniaturization.

**Methods**

*Fabrication:* The photonic-crystal device consists of a square lattice of holes etched into a SOI wafer. The lattice constant *a* is 860 nm, and the hole radius is 0.174*a* (150 nm). The thickness of the silicon device layer is 250 nm. The buried oxide layer is 3 μm thick, and the silicon substrate is 600 μm thick. To fabricate the device, a layer of PMMA 4% in anisole was spin coated onto an SOI wafer. Electron beam lithography at 30 kV was used to write the photonic-crystal pattern into the resist. A modified Bosch process was used to transfer the pattern into the Si device layer using a mixture of $SF_6$ and $C_4F_8$ gases. Plasma-enhanced chemical vapor deposition was used to deposit a 193 nm layer of $SiN_x$ onto the polished, back surface of the sample to reduce unwanted reflections.

The photonic-crystal sample is mounted on a circular glass slide and inserted in a rotary stage. The sample is covered with a solution of 520 nm diameter polystyrene particles (700 μL of Thermo Scientific Fluoro-Max R520 particles in suspension, diluted in 60 mL of deionized water). 600 μL of 1% Triton-X was added to the solution to minimize particle stiction to the sample surface. On a second glass slide, an open-topped microfluidic chamber (4 mm x 4 mm) was fabricated in a 5-μm layer of PDMS using photolithography. The PDMS chamber was pressed on to the sample and sealed inside the rotary stage.

*Optical setup:* A Santec tunable laser with a tuning range from 1500 nm to 1620 nm was connected to an erbium-doped fiber amplifier and a tunable bandpass filter with a width of 1 nm. An adjustable, neutral-density filter and polarization-control optics were used to control the power and polarization of the beam. A fiber-to-free-space collimator directed the beam to the entrance aperture of a 20X objective (NA = 0.40), which focused the beam onto the top surface of the sample from the back side. A second 20X objective was used in conjunction with a beam splitter to collect the light from the top surface and to image the particle motion on a CMOS camera. Prior to each self-assembly experiment, the transmission was measured in cross-polarization mode. Polarizers before and after the sample were oriented at 90 degrees from one another, and the wavelength of the transmission peak was identified. This laser was then tuned to the peak wavelength to carry out light-assisted self assembly.

*Stiffness analysis:* After the assembly of a cluster, we recorded videos with a fixed exposure time of 33 ms, the fastest value available in our experimental set-up. Typical videos were 600 or more frames in length. Particle motion was analyzed using MATLAB algorithms written by Blair and Dufresne (http://physics.georgetown.edu/matlab/), which are modified versions of IDL routines written by Crocker and Grier.

The blue ellipses in Figure 4(a) are obtained by fitting the data in each unit cell. The direction and relative lengths of the major and minor axis are determined from the eigenvectors and eigenvalues of the scaled covariance matrix. The ellipse is drawn to represent a 95% confidence interval: positions outside the ellipse will be observed only 5% of the time by chance if the underlying distribution is Gaussian.

The measured variances were corrected for motion blur due to the finite integration time of the camera.[29] We observed that the variance as a function of position within the cluster had a 2D Gaussian distribution. This is to be expected due to the spatial variation in intensity in the incident beam. We confirmed via 3D-FDTD simulations of a finite-size structure with Gaussian beam excitation that the field intensity above the slab has a Gaussian envelope. The stiffness values shown in Figure 4 are normalized to the local power at each trapping site, as determined by fitting the experimentally-measured variances to a 2D spatial Gaussian.

We directly measured the diffusion coefficient by first assembling a cluster, then blocking the beam and performing a linear regression fit to the subsequent diffusion as a function of time.[30] A value of 0.56 µm$^2$/s was obtained.

*Force measurement:* After assembling a cluster, we increased the flow speed until all particles were released. We observed the release of the last, trapped particle at a fluid velocity of 30 µm/s for a laser power of 64 mW. In the low Reynolds number regime, the particle velocity $v$ is related to the external force by the equation $F = 6\pi\eta r v$, where $\eta$ is the viscosity of the medium, and $r$ is the radius of the particle.

*Force and potential calculation:* For each particle position within the unit cell, we performed an FDTD simulation to calculate the electromagnetic fields and obtained the optical force from an integral of the Maxwell Stress Tensor over a box just large enough to include the particle. The *z*-dependence of the force inside the hole was obtained from an exponential fit to the field intensity above the slab. The power threshold for stable trapping was determined experimentally from the size of the stable cluster (16 unit cells, or 13.8 µm, in diameter). The Gaussian distribution of energy in the mode, obtained from a fit of the variance measurements, had a FWHM of 15 µm. Given the input power of 64 mW we find a peak intensity of 240 µW per unit cell, and a trapping threshold intensity at the edge of the cluster of 134 µW per unit cell.

**Acknowledgments**

The authors thank Mia Ferrera Wiesenthal for rendering the schematic in Figure 1. This project was funded by an Army Research Office PECASE Award under grant no. 56801-MS-PCS. Computation for work described in this paper was supported by the University of Southern California Center for High-Performance Computing and Communications. The authors thank Ningfeng Huang, Jing Ma and Chenxi Lin for fruitful discussions.

**Author Contributions**

E. J. and L. J. M. fabricated the devices and performed the experiments. C. A. M. performed the simulations and the data analysis. E. J. and M. L. P. wrote the manuscript with contributions from all authors. M. L. P. conceived of the experiment.

**Competing Financial Interests**

The authors declare no competing financial interests.

**Figure Legends**

Figure 1 – Schematic of light-assisted, templated self assembly (LATS). Incident light from below excites a guided-resonance mode of a photonic-crystal slab, giving rise to optical forces on nanoparticles in solution. Under the influence of the forces, the nanoparticles self assemble into regular, crystalline patterns.

Figure 2 – Square lattice photonic crystal. (a) SEM image of photonic-crystal slab. The scale bar in the inset is 1 μm. (b) 3D FDTD simulation of the magnetic field ($H_z$) for a normally-incident, *x*-polarized plane wave. Circles represent the positions of holes; four unit cells are shown. (c) Measured transmission spectrum (log scale).

Figure 3 – Light-assisted, templated self assembly of 520-nm diameter particles above a photonic-crystal slab. The square lattice of the slab is visible in the background, oriented at 45° with respect to the camera. (a-c) Sequential snapshots taken with the light beam on. (d) Snapshot taken after the beam is turned off.

Figure 4 – Trap stiffness for incident, *x*-polarized light. (a) Particle positions (red dots) extracted from a 20-second video. Blue ellipses represent two standard deviations in position. (b) Histogram of stiffness values in the direction parallel to the polarization of the incident light. (c) Stiffness in the direction perpendicular to the incident polarization.

Figure 5 – Calculated optical forces and potentials. (a) Force map in one unit cell. The black circle represents the position of a hole, and the color bar represents the vertical optical force in dimensionless units of Fc/P, where $c$ is the speed of light, and $P$ is the incident optical power. A negative force indicates attraction toward the slab. The arrows represent the magnitude of the lateral optical force. The length of the white arrow at the bottom center of the figure corresponds to the maximum in-plane value of 1.6. (b) The potential through the center of the hole along the $x$-direction. The green dashed lines indicate the position of the hole. (c) The potential through the center of the hole along the $y$-direction.

**Table 1 – Polarization dependence of trap stiffness**

| angle | $\kappa_x$ (pN nm$^{-1}$ W$^{-1}$) | $\sigma_x$ | $\kappa_y$ (pN nm$^{-1}$ W$^{-1}$) | $\sigma_y$ |
|---|---|---|---|---|
| **0°** | 1.48 | 0.35 | 2.25 | 0.48 |
| **45°** | 1.88 | 0.52 | 1.79 | 0.46 |
| **90°** | 2.08 | 0.40 | 1.39 | 0.29 |

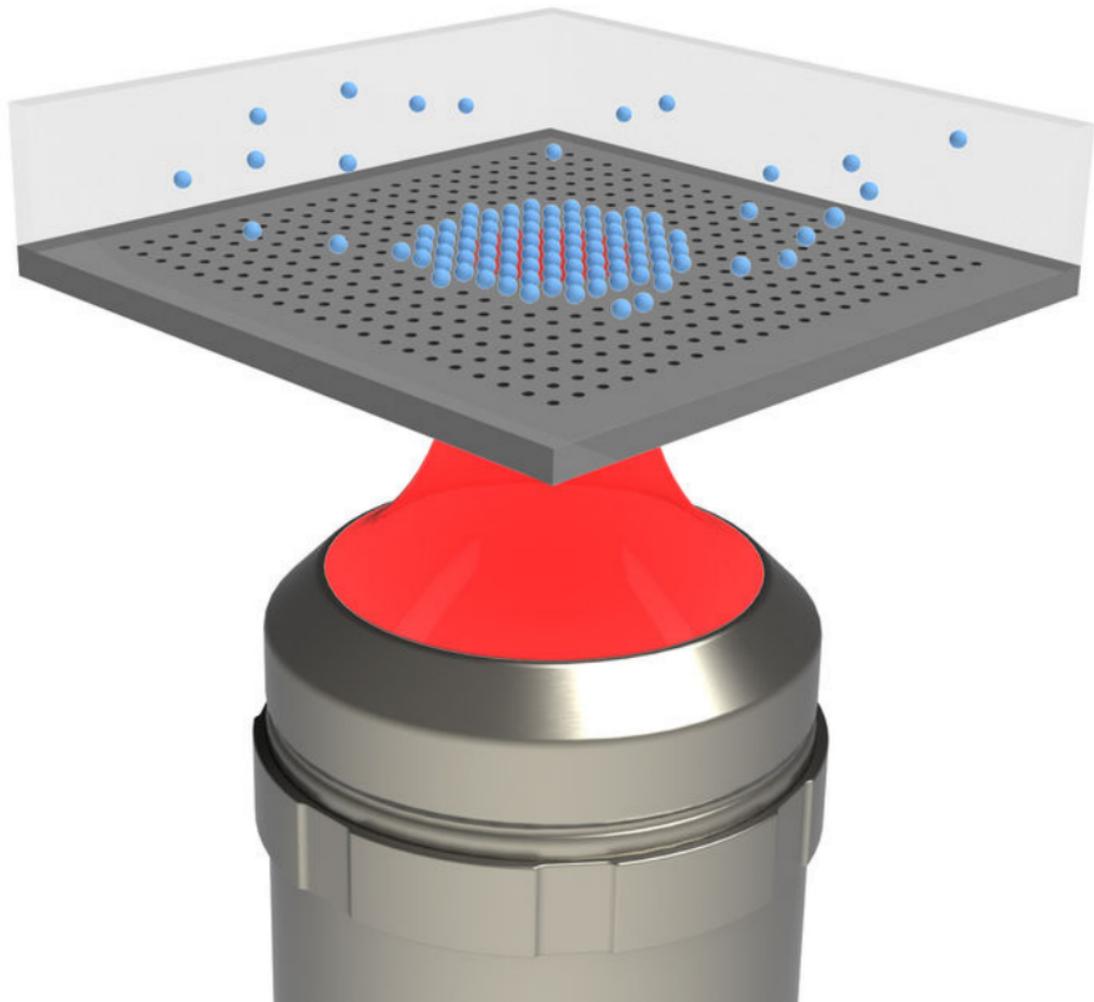

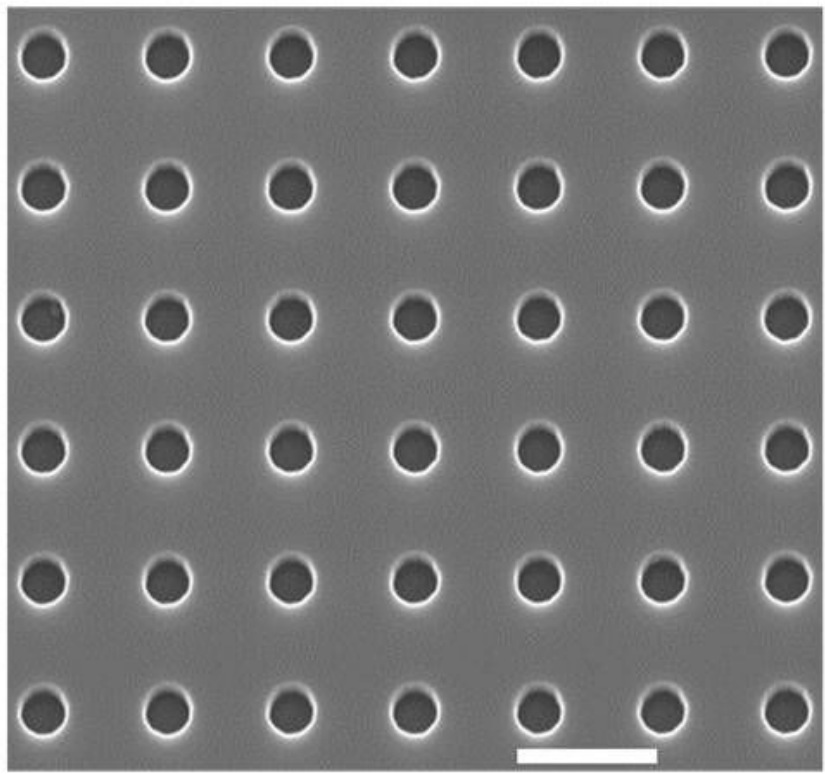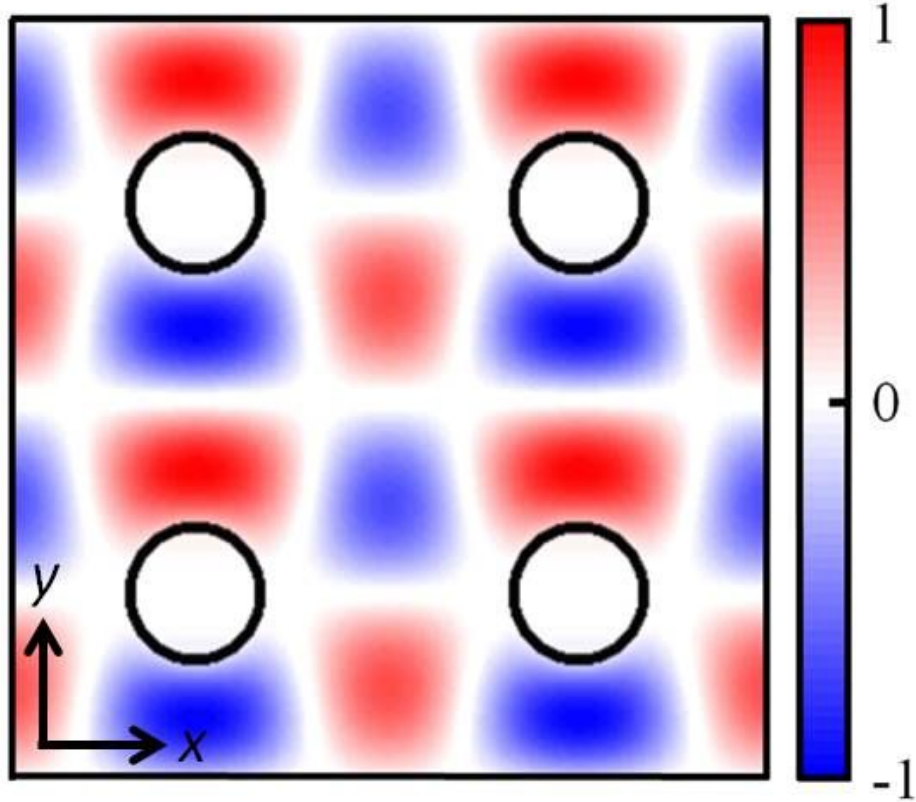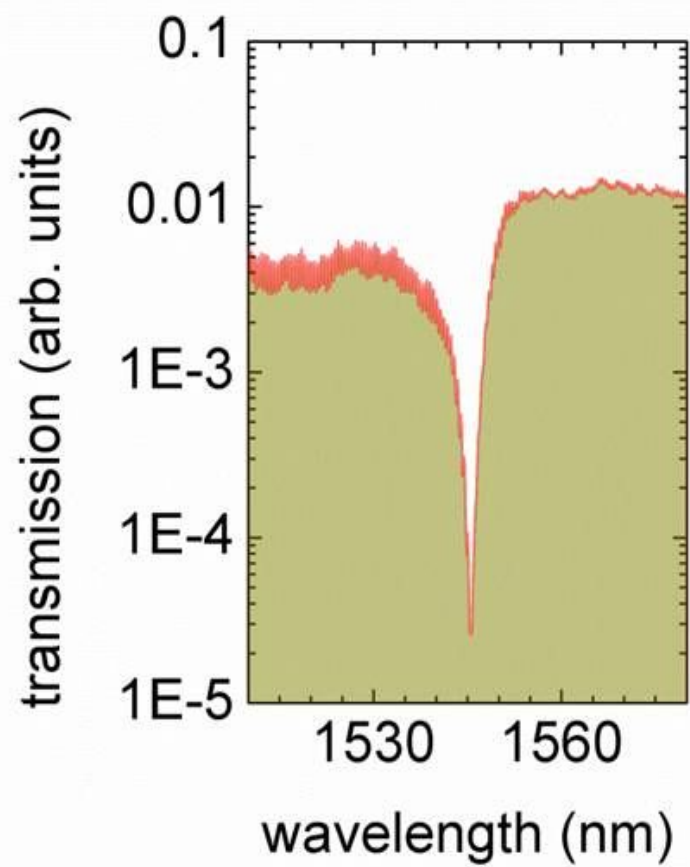

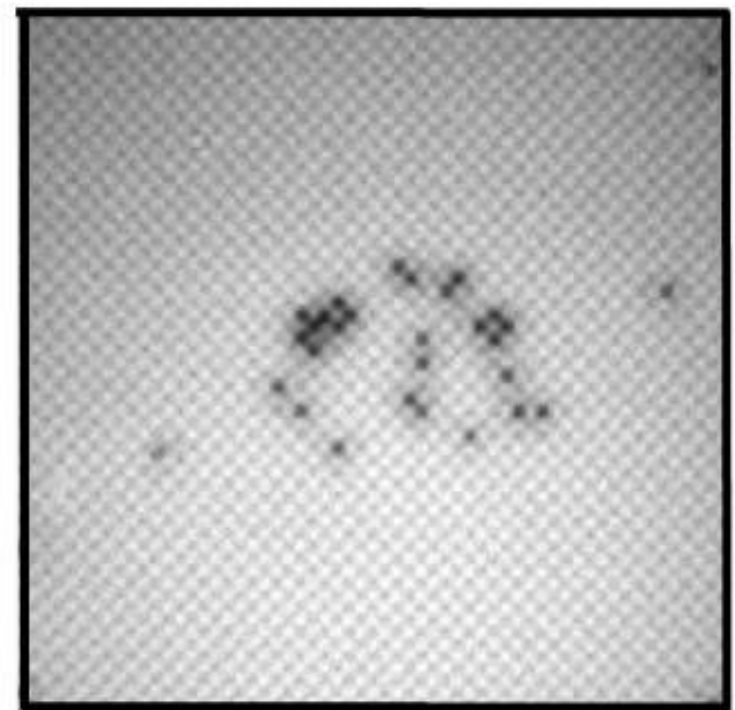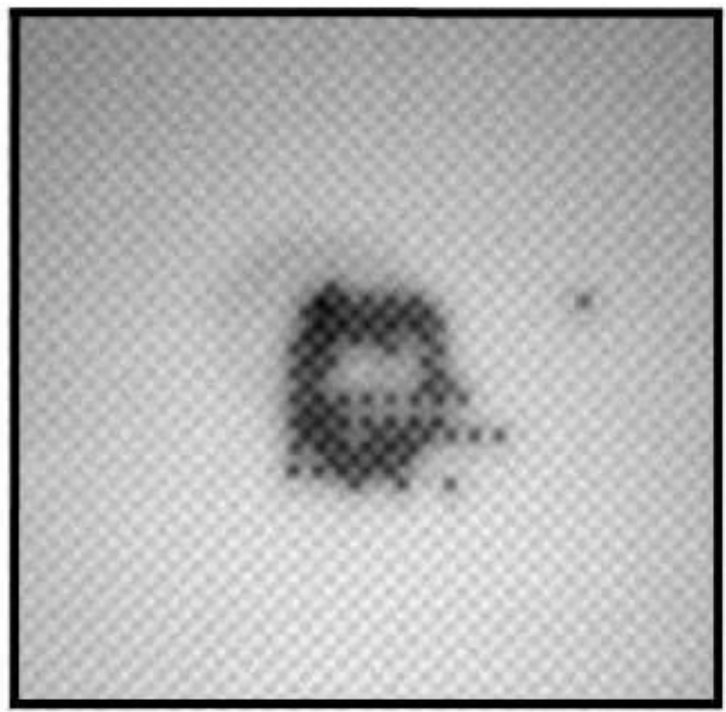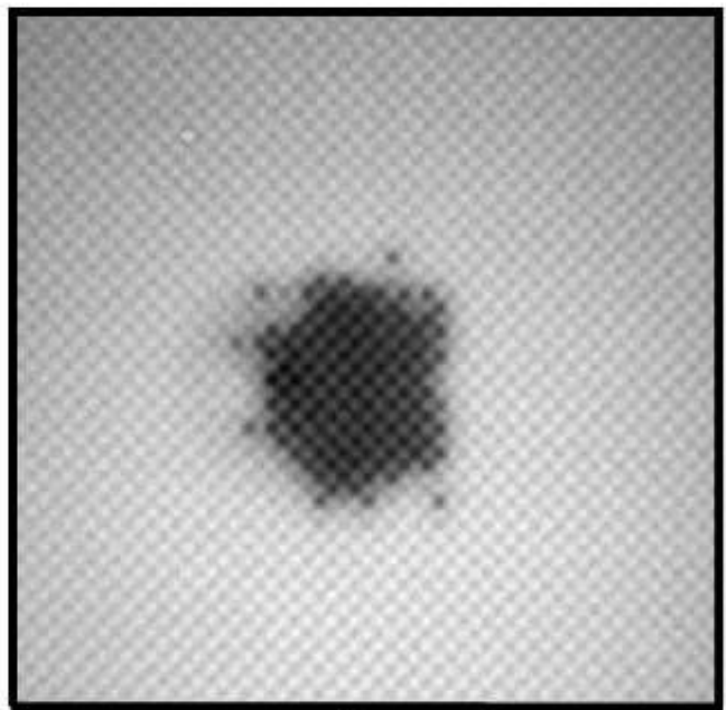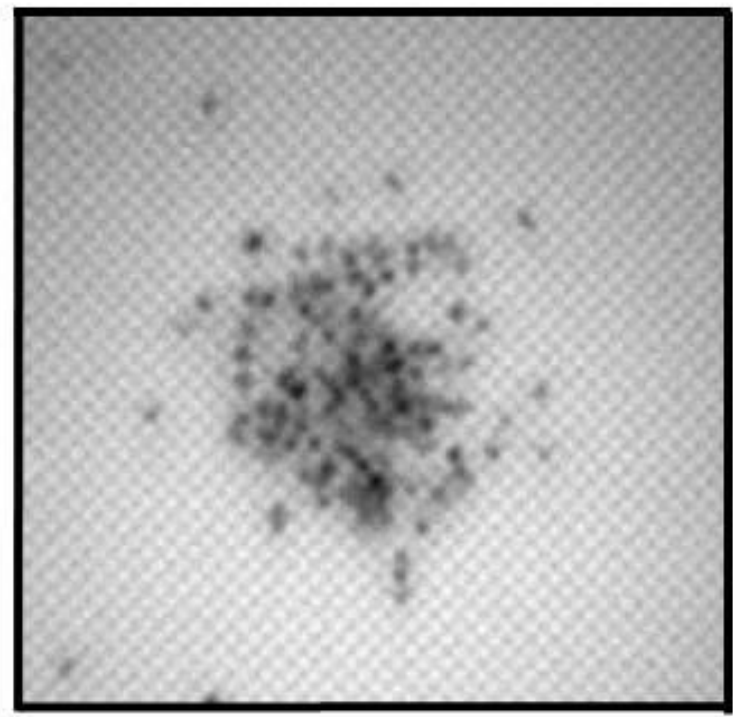

assembly → release

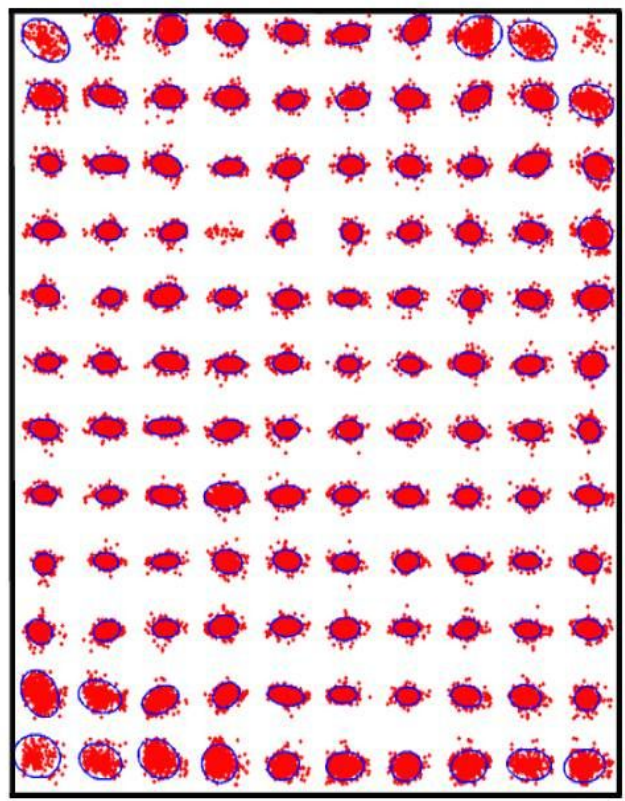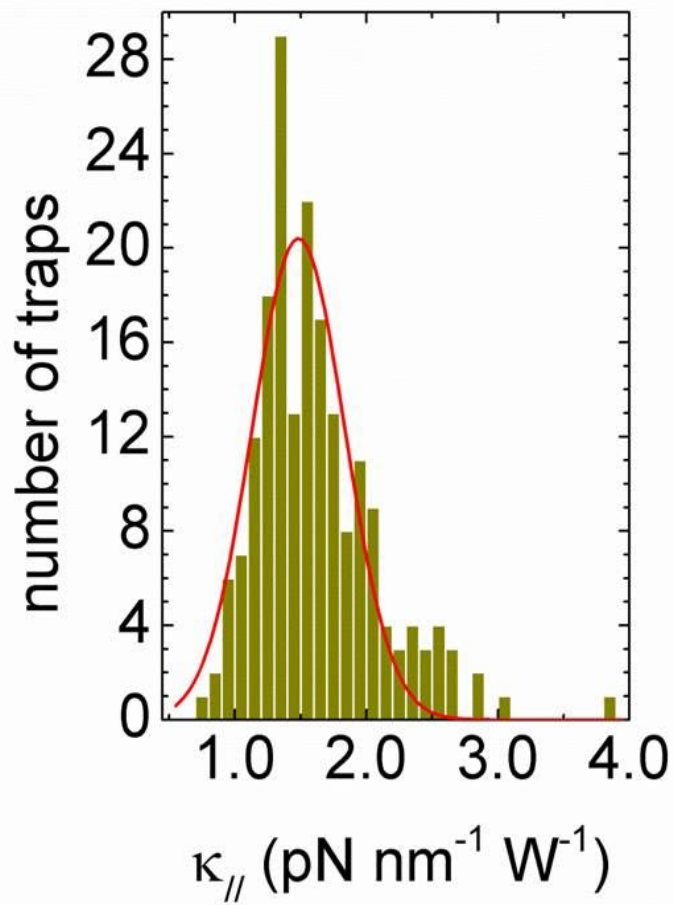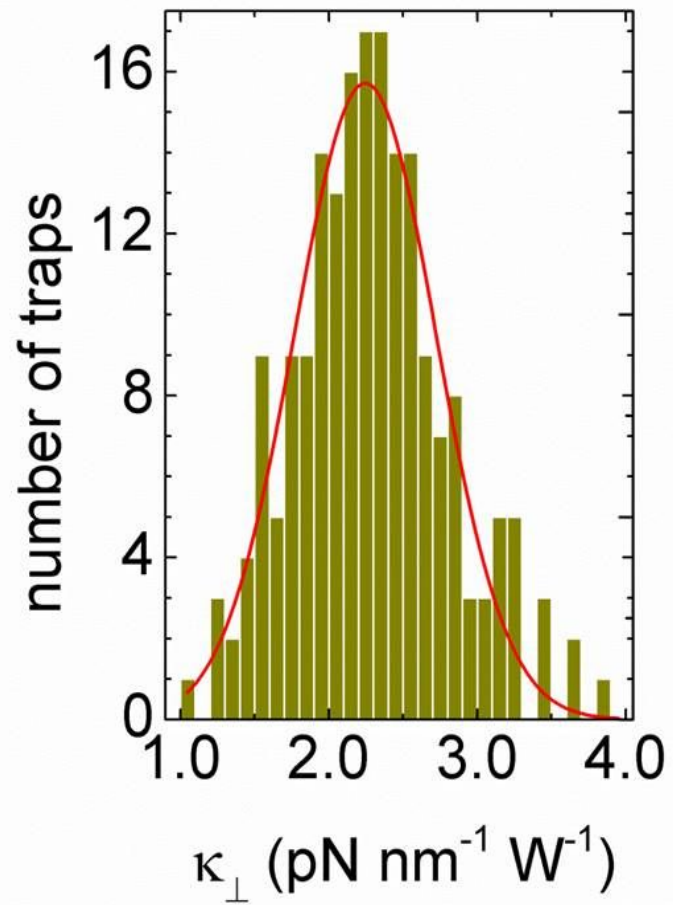

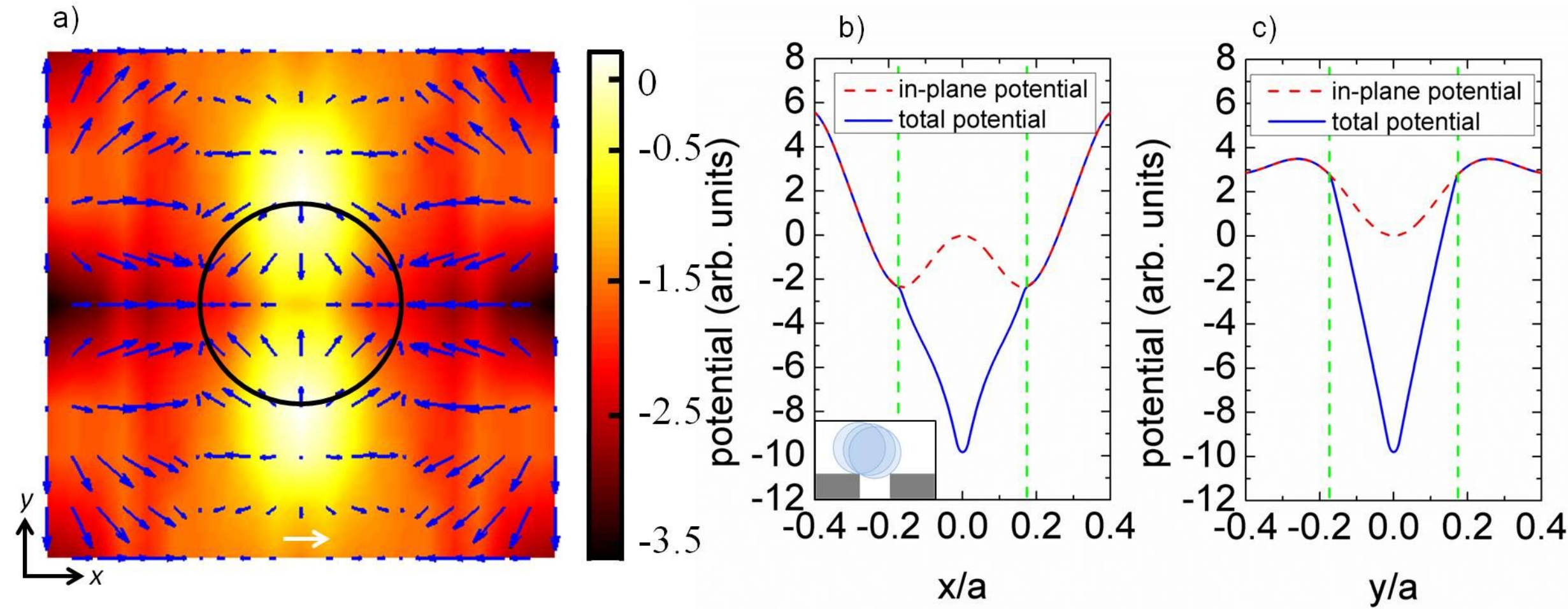